\documentclass[10pt,twoside]{article}
\usepackage{amsfonts}
\usepackage{fancyhdr}
\usepackage{titlesec}
\usepackage{cite}
\usepackage{ifthen}
\usepackage{graphicx}

\titleformat{\section}{\centering\large\bfseries}{\S\arabic{section}}{1em}{}
\newboolean{first}
\setboolean{first}{true}

\begin{document}

\setlength\belowdisplayskip{2pt}
\setlength\belowdisplayshortskip{0pt}

\title{\textbf{\Large The analysis on the single particle model of CDW}}
\author{Lian-Gang Li, Yong-Feng Ruan \\
{\small \emph{Department of Physics, School of Sciences, Tianjin
University, Tianjin 300072, China}}\date{}} \maketitle

\footnote{
Lian-Gang Li: Ph. Doctor, \emph{e-mail}: liliank@tju.edu.cn \\
Yong-Feng Ruan: Professor, \emph{e-mail}: Ruanyf@tju.edu.cn}

\begin{center}
\begin{minipage}{135mm}
{\bf \small Abstract}.\hskip 2mm {\small Gr\"{u}ner put forward a
single particle model of charge-density wave, which is a typical
nonlinear differential equation, and also a mathematical model of
pendulum. This Letter analyzes the solution of equation by the
rotated vector fields theory, providing the relation between the
applied field $E$ and the periodic solution, and the conclusion that
the critical value of $E$ for the periodic solution is fixed in the
over-damped situation. With these conclusions, it derives the
formulae of nonlinear conductivity, narrow-band noise, which are
consistent with the empirical ones given by Fleming.}

\vskip 3mm {\small \textit{PACS}:}\hskip 2mm {\small 71.45.Ir;
02.30.Hq} \vskip 3mm {\small \emph{Keyword}:}\hskip 2mm {\small
charge-density wave (CDW); pendulum; rotated vector field; periodic
solution; narrow-band noise; Gr\"{u}ner's equation}
\end{minipage}
\end{center}

\vskip 6mm

\setcounter{section}{1} \setcounter{equation}{0}
\renewcommand{\theequation}
{\arabic{equation}} \setcounter{figure}{0}\renewcommand{\thefigure}{%
\arabic{figure}}

\begin{flushleft}
{\large \textbf{1.\hspace*{2mm}Introduction}}
\end{flushleft}

Charge-density waves (CDW) is a special phenomenon of electronic
condensate in the low dimensional materials and observed in the High
Temperature Superconductor materials as well\cite{kim}. Many special
features associated with CDW, for example, the nonlinear
conductivity, narrow-band noise, have been attracting great
attention\cite{Lee74,Fuku,Lee79,Moro}. Describing the nonlinear
conductivity and narrow-band noise, Gr\"{u}ner and his collaborators
proposed a single particle model\cite{Grun}, of which the
equation is%
\begin{equation}
\frac{\mathrm{d}^{2}\phi }{\mathrm{d}t^{2}}+\Gamma \frac{\mathrm{d}\phi }{%
\mathrm{d}t}+\sin \phi =\beta ,  \label{eq:f1}
\end{equation}%
where $\beta =E/E_{0}$ with $E_{0}=( \lambda /2\pi
)(m\omega _{0}^{2}/e)$ is constant and $E$ is an electric field applied, $%
\Gamma $ is the friction coefficient. Equation (\ref{eq:f1}) is called Gr\"{u%
}ner's equation in the CDW theory as well as is the mathematical
model of driven, damped pendulum. This is a typical nonlinear
differential equation which is non-integrable. In the over-damped
situation, Gr\"{u}ner gave an approximate solution with neglect of
the inertial term\cite{Grun}. The derived conductivity, i.e. $\sigma
(E)=\sigma _{0}\sqrt{E^{2}-E_{0}^{2}}$, has an excellent behavior
served as the exponent law given in Fleming's paper in the region of
high field, but this neglect obviously loses some characteristics of
original equation, such that it leads to a divergence of the
derivative at the threshold field, a intrinsic difference from the
experiment result\cite{Flem}.

There are two approaches in the research of nonlinear differential equation.
The first one is a classical method which attempts to express the solutions
of differential equations into primary functions named closing, or into
power series form. This method is difficult to develop further. The second
is so-called "the vector field theory of differential equation", originated
by Henri Poincar\'{e}\cite{Poin}, which treats the solutions as integral
curves in the phase space and derives the qualitative properties of the
solution geometrically. Many scientists have studied in this theory's area,
such as, G. F. Duff proposed the rotated vector fields in 1953\cite{Duff},
subsequently, G. Seifert \emph{et al.} developed it to the general rotated
vector fields\cite{Seif,Chen,Ma}.

With general rotated vector fields, this Letter provides some
conclusions on
the solutions of equation (\ref{eq:f1}), that is, the relation between $%
\beta $ and the periodic solution, and that, with $\Gamma $ big enough (over
damping), the critical value $\beta _{0}$ will remain unchanged, i.e., $%
\beta _{0}\equiv 1$. Combining these conclusions with the multiple segments
model advanced by Portis\cite{Port}, it derives the same formulae as the
empirical ones given by Fleming\cite{Flem}.

\begin{flushleft}
{\large \textbf{2.\hspace*{2mm}The periodic solution of the equation}}
\end{flushleft}

On the $\phi -z$ phase plane, Equation (\ref{eq:f1}) takes the form of
vector field:
\begin{equation}
\left\{
\begin{array}{ll}
\frac{\mathrm{d}\phi }{\mathrm{d}t}=z &  \\
\frac{\mathrm{d}z}{\mathrm{d}t}=\beta -\sin \phi -\Gamma z &
\end{array}%
\right. ,  \label{eq:f2}
\end{equation}%
where $\Gamma >0$, $\beta \geq 0$, whose solutions correspond to the
trajectories on the plane. Each trajectory, has a direction running
as time goes forward, to which the tangents become vectors which
constitute the vector field. When the vectors are rotating owing to
the change of some parameter of equation, they constitute the
rotated vector field. A point satisfying $\frac{d\phi}{dt}=0$ and
$\frac{dz}{dt}=0$ is called a "singularity". A field, in which the
singularities happen to move accompanying the rotation of vectors,
is called the general rotated vector field. In the theory of
differential equation, owing to the uniqueness of the solution, the
trajectories do not intersect each other except at
singularities\cite{Perk}. This is very useful property in the
following analysis.

\begin{figure}[h]
\centering
\includegraphics[width=3.20in, height=1.60in]{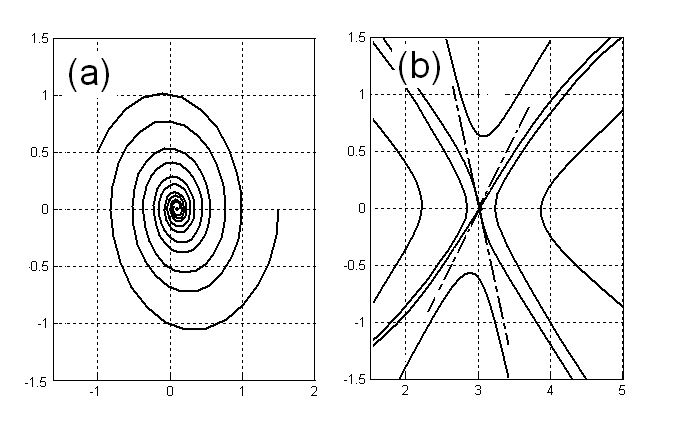}
\caption{The plot of singularities: (a) focus, (b) saddle}
\label{fig:p1}
\end{figure}

While $\beta >1$, there exist no singularities. While $0\leq \beta \leq 1$,
there exist singularities on the $\Phi $-axis with coordinates denoted by $%
(\phi _{n},0)$ where
\begin{equation}
\left\{
\begin{array}{ll}
\phi _{0}=\arcsin \beta &  \\
\phi _{n}=n\pi +(-1)^{n}\phi _{0} &
\end{array}%
\right. ,\quad \quad n\in \mathbb{Z},  \label{eq:f3}
\end{equation}%
which are categorized into two classes, one denoted by $A_{k}(\phi
_{2k-1},0) $, the other denoted by $B_{k}(\phi _{2k},0)$, $k\in \mathbb{Z}$.
While $\Gamma >0$, $B_{k}(\phi _{2k},0)$ is called a "focus", which is
stable, nearby which all the trajectories spiral into it as time goes
forward, as shown in Figure \ref{fig:p1} (a). While $\Gamma =0$, it is
called a "center", which is not the subject of the current analysis. $%
A_{k}(\phi _{2k-1},0)$ is called a "saddle", nearby which there are four
trajectories approaching it along two separatrices, as shown in Figure \ref%
{fig:p1} (b). These singularities are key to analyze the equation (\ref%
{eq:f2}).

\begin{figure}[h]
\centering
\includegraphics[width=1.40in, height=1.60in]{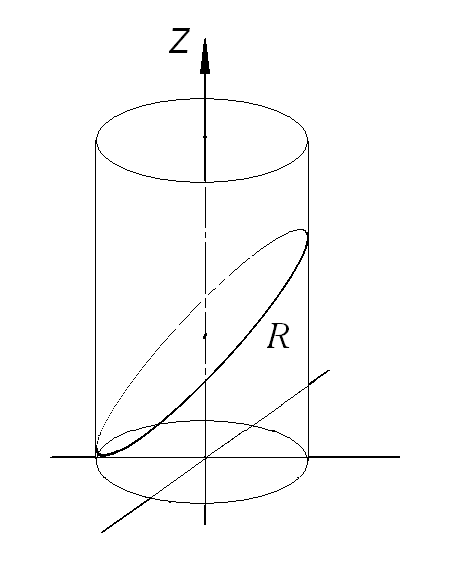}
\caption{Cycle on the cylindrical surface} \label{fig:p2}
\end{figure}

For convenience, let $W(\phi ,z)=z$, $Q(\phi ,z)=\beta -sin\phi -\Gamma z$.
The angle between the vector and the $\Phi $-axis is denoted by $\theta $,
where
\begin{equation}
\theta =\tan ^{-1}\frac{Q(\phi ,z)}{W(\phi ,z)}.  \label{eq:f4}
\end{equation}%
Clearly, the vector field constructs a rotated vector field when parameter $%
\Gamma $ or $\beta $ changes. It is different that the singularities
do not move as parameter $\Gamma $ changes but move as parameter
$\beta $ does, so it is a general rotated vector field with respect
to $\beta $. The distribution of vectors is determined by $W(\phi
,z) $ and $Q(\phi ,z)$, both with a period of $2\pi $ along the
$\Phi $-axis, so the vectors distribute with the same period and the
plane can be rolled up into a cylindrical surface, that is called a
cylindrical system in mathematics. Thus it is enough for analysis
that we just discuss the interval $[-\pi -\phi _{0},\pi -\phi _{0}]$
without special declaiming, whereas the others can be extended with
periods along the $\Phi $-axis. In the cylindrical system, the
trajectory, which connects successive periodic points, will become a
cycle on the cylindrical surface, as shown in Figure \ref{fig:p2},
which is a periodic solution. It is found that equation
(\ref{eq:f2}) just possesses such a periodic solution in some
condition.

\begin{figure}[h]
\centering
\includegraphics[width=2.80in, height=1.79in]{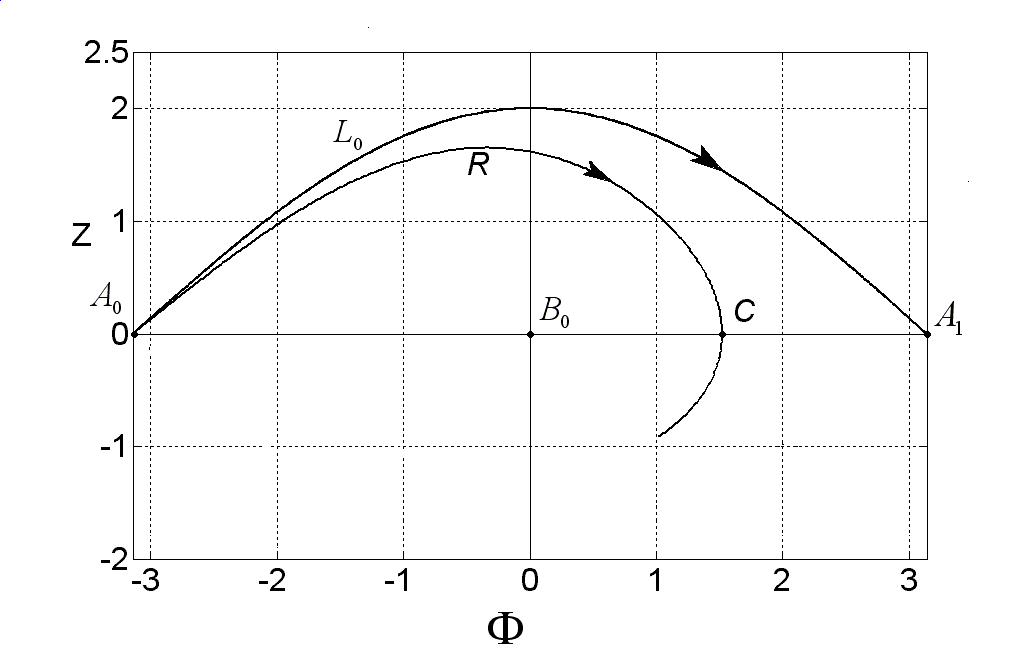}
\caption{Trajectories as $\protect\beta =0$}
\label{fig:p3}
\end{figure}

It is very difficult to tackle the equation (\ref{eq:f2}) directly, but in
the case where $\Gamma =0$ and $\beta =0$, it offers a simple equation,
i.e., ${\displaystyle\frac{\mathrm{d}^{2}\phi }{\mathrm{d}t^{2}}+\sin \phi =0%
}${, }which is called sine-Gordon's equation, for which it is easy
to give a solution,
\begin{equation}
z=\sqrt{2(\cos \phi +1)}.  \label{eq:f5}
\end{equation}%
Its corresponding trajectory $L_{0}$ connects saddle point $A_{0}$ with
saddle point $A_{1}$, as shown in Figure \ref{fig:p3}. Point $A_{1}$ is a
periodic point of $A_{0}$, so trajectory $L_{0}$ is a cycle on the
cylindrical surface, a periodic solution. As connecting two successive
saddles, it is called a critical periodic solution in mathematics.

For which, let $\Gamma $ increase from zero to a positive, by the analysis
of rotated vector field, it is proved that trajectory $L_{0}$ will roll down
and become trajectory $R$ as shown in Figure \ref{fig:p3}. While $\Gamma >0$%
, let $\beta $ increase from zero to a positive value, it is proved
that trajectory $R$ will stretch up and singularity $A_{1}$ moves
toward left as well, such that trajectory $R$ enters singularity
$A_{1}$ again at a certain positive value of $\beta $ which is no
greater than $1$\cite{Arxiv}. Since point $A_{1}$ is the successive
periodic point of $A_{0}$, trajectory $R$ connecting these two
saddles will become a cycle on the cylindrical surface, a critical
periodic solution, as shown in Figure \ref{fig:p2}.

In other words, equation (\ref{eq:f2}) must possess a value $\beta
_{0}$ for arbitrary $\Gamma >0$, while $\beta =\beta _{0}$, equation
(\ref{eq:f2}) has
a critical periodic solution which connects saddle $A_{0}$ with saddle $%
A_{1} $. By the analysis of rotated vector field, it can also be proved
that, while $\beta \geq \beta _{0}$, equation (\ref{eq:f2}) possesses a
unique and positive periodic solution, which is stable as well; while $\beta
<\beta _{0} $, equation (\ref{eq:f2}) possesses no periodic solution.
Through analysis, it shows that $\beta _{0}$ is unique for a given $\Gamma
>0 $.

According to equation (\ref{eq:f2}), it is easy to obtain%
\begin{equation}
\frac{\mathrm{d}z}{\mathrm{d}\phi }=\frac{\beta -\sin \phi }{z}-\Gamma .
\label{eq:f6}
\end{equation}%
On the base of equality (\ref{eq:f6}), it is easy to obtain Theorem
1 as follows.

\noindent \textbf{Theorem 1.} \textit{If a periodic solution }$z=z(\phi )$
\textit{exists for the equation (\ref{eq:f2})}, i.e., $z(\phi )=z(\phi +2\pi
)$, $\phi \in (-\infty ,+\infty )$, \textit{then it must satisfy}%
\begin{equation}
\int_{\phi }^{\phi +2\pi }z(\phi )\,\mathrm{d}\phi =\frac{2\pi \beta }{%
\Gamma }.  \label{eq:f7}
\end{equation}

The proof of Theorem 1 is very simple. According to equality
(\ref{eq:f6}),
integrating from $\phi $ to $\phi +2\pi $ and noticing the periodicity of $%
z(\phi )$, it is easy to obtained equality (\ref{eq:f7}).

According to Theorem 1, it is easy to prove that the periodic
solution is unique and positive. In the proof, it just need to
notice the property that the trajectories do not intersect each
others.

According to the following reason, it is easy to prove that the periodic
solution is stable. The periodic solution is a cycle on the cylindrical
surface, of which the characteristic exponent is $-\Gamma <0$, therefore it
is a stable limit-cycle\cite{Perk}, a stable periodic solution.

\begin{figure}[h]
\centering
\includegraphics[width=2.90in, height=1.90in]{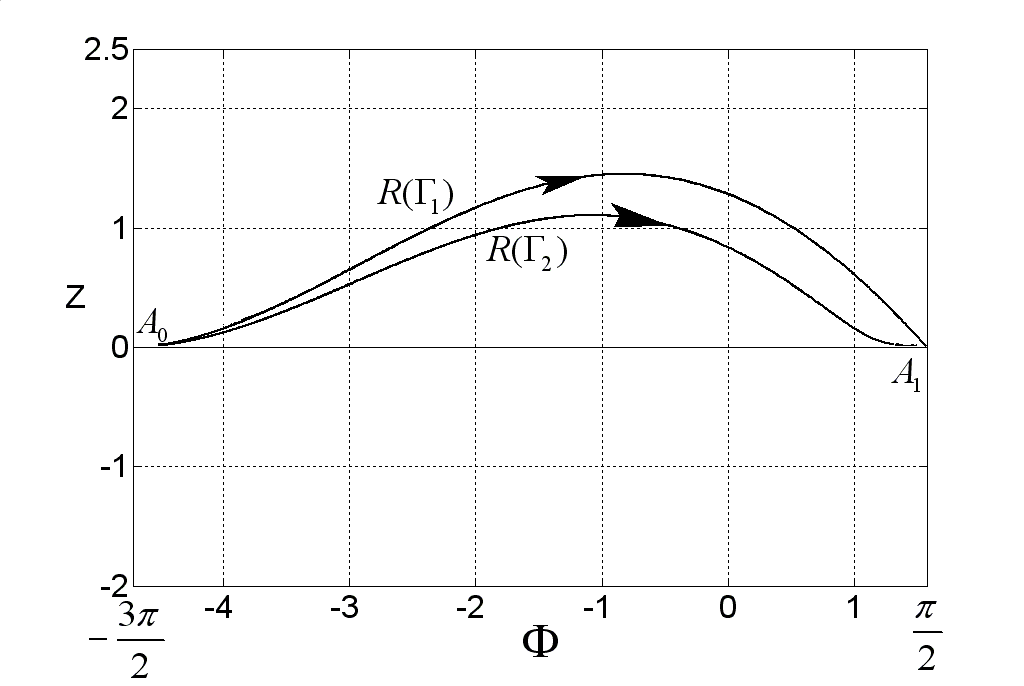}
\caption{The critical periodic trajectories as $\Gamma _{2}>\Gamma
_{1}$} \label{fig:p4}
\end{figure}

Supposing $\beta _{0}(\Gamma _{1})=1$ and $\Gamma _{2}>\Gamma _{1}$, it can
be proved that, $\beta _{0}(\Gamma _{2})=1$, that is to say, the critical
value $\beta _{0}$ equals $1$ all the time if $\Gamma $ is bigger enough.
Interestingly, the slope of $R(\Gamma _{2})$ at point $A_{1}$ is always $0$,
as shown in Figure \ref{fig:p4}

\begin{flushleft}
{\large \textbf{3.\hspace*{2mm}Interpretation of the periodic solution in
physics}}
\end{flushleft}

\begin{figure}[h]
\centering
\includegraphics[width=2.90in, height=1.51in]{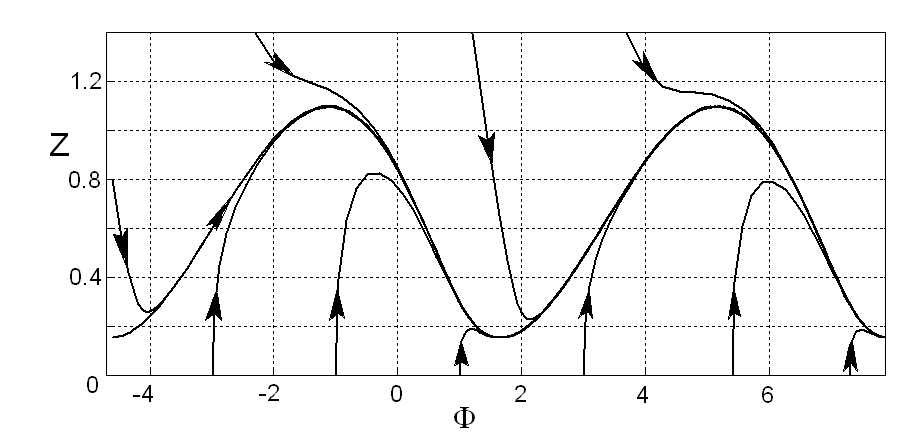}
\caption{A stable, positive periodic solution}
\label{fig:p5}
\end{figure}

According to the conclusion above, it follows that while $\beta \geq \beta
_{0}$, Equation (\ref{eq:f2}) must possess a unique, stable and positive
periodic solution, where a stable trajectory means that all those nearby
will approach it as time goes forward. It indicates that, whatever the
initial state it is, the final state will be in a periodic motion toward the
positive direction, as shown in Figure \ref{fig:p5}.

\begin{figure}[h]
\centering
\includegraphics[width=2.90in, height=1.51in]{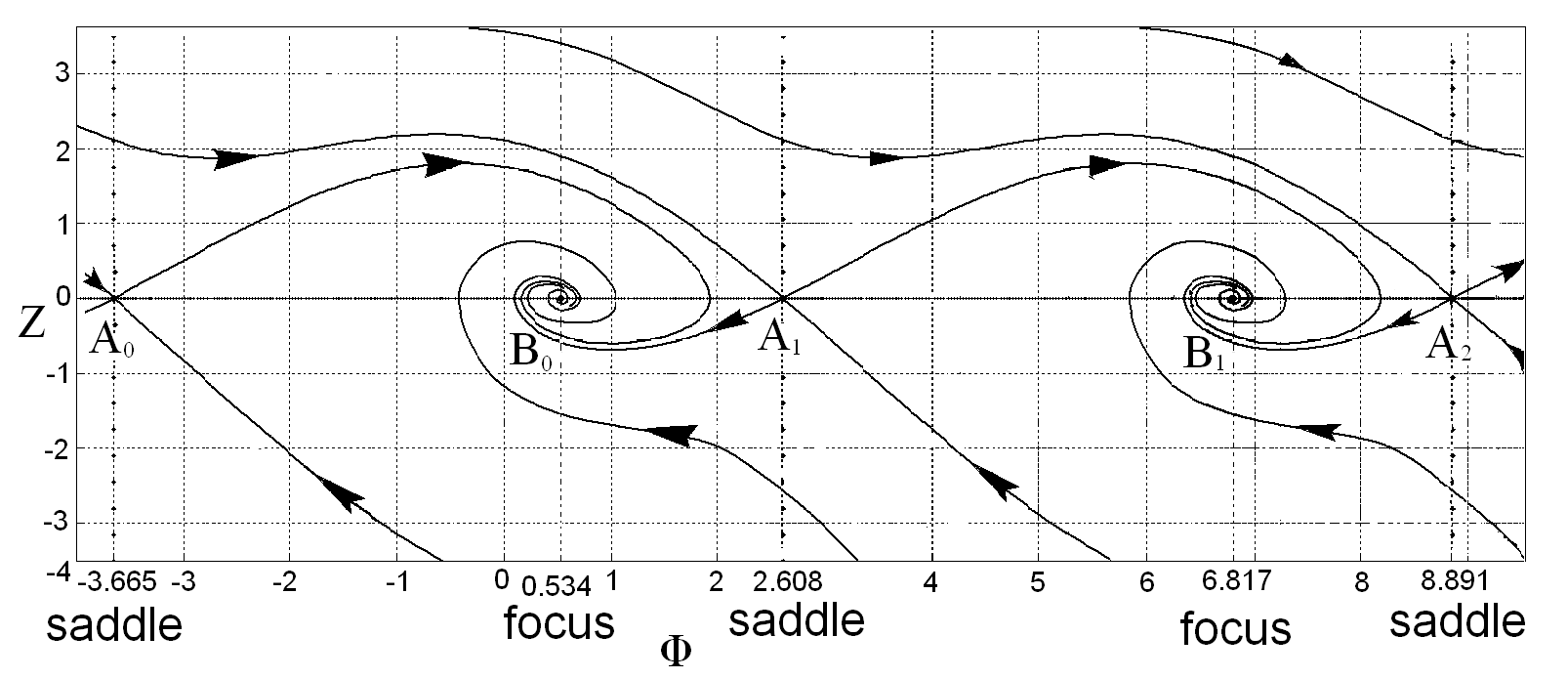}
\caption{Nonperiodic trajectories}
\label{fig:p6}
\end{figure}

While $\beta <\beta _{0}$, Equation (\ref{eq:f2}) possesses no periodic
solution. In fact, all trajectories on the phase plane will approach the
saddles or foci. If there are some disturbances, they will deviate from the
saddles, which are unstable; finally, all the trajectories will end at the
foci, which are stable, as shown in Figure \ref{fig:p6}. In the physics,
whatever the initial state it is, the system will stay at equilibrium points
(foci) in the stable state.

Supposing a minimum value $\Gamma _{\min }$ such that $\beta _{0}(\Gamma
_{\min })=1$, as above it follows that, while $\Gamma >\Gamma _{\min }$, the
critical value $\beta _{0}$ will be $1$ all the time. M. Urabe has given the
definite value of it by numerical calculations\cite{Urab}, that is, $\Gamma
_{\min }\approx 1.193$. According to the experimental values given in the
article\cite{Port}, $\omega _{0}\approx 2\pi \times 210(MHz)$, $\tau
=1.3\times 10^{-10}(\sec )$, $\Gamma =(\omega _{0}\tau )^{-1}$, it derives $%
\Gamma =5.83$. Clearly, in the practical case, $\Gamma $ is greater than $%
\Gamma _{\min }$ so far so that $\beta _{0}=1$. Because $\beta
=E/E_{0}$, it follows that, while $E\geq E_{0}$, Equation
(\ref{eq:f1})
possesses a unique periodic solution which satisfies Theorem 1, i.e.,%
\begin{equation}
\frac{2\pi E}{\Gamma E_{0}}=\int_{0}^{2\pi }\frac{\mathrm{d}\phi }{\mathrm{d}%
t}\,\mathrm{d}\phi .  \label{eq:f8}
\end{equation}%
Let $\frac{\mathrm{d}\phi }{\mathrm{d}t}=\left( \frac{\mathrm{d}\phi }{%
\mathrm{d}t}\right) _{c}+\left( \frac{\mathrm{d}\phi }{\mathrm{d}t}\right)
_{a}$, where%
\begin{equation}
\left( \frac{\mathrm{d}\phi }{\mathrm{d}t}\right) _{c}=\frac{E-E_{0}}{\Gamma
E_{0}},  \label{eq:f9}
\end{equation}%
\begin{equation}
\frac{1}{2\pi }\int_{0}^{2\pi }\left( \frac{\mathrm{d}\phi }{\mathrm{d}t}%
\right) _{a}\,\mathrm{d}\phi =\frac{1}{\Gamma },  \label{eq:f10}
\end{equation}%
$\left( \frac{\mathrm{d}\phi }{\mathrm{d}t}\right) _{c}$ is the uniform
velocity of moving as an ensemble and contributing to the direct current
conductivity, $\left( \frac{\mathrm{d}\phi }{\mathrm{d}t}\right) _{a}$ is
the modulated part producing the narrow-band noise. As the difference of
velocity between segments results the pressing and stretching each other,
the constant component in $\left( \frac{\mathrm{d}\phi }{\mathrm{d}t}\right)
_{a}$, i.e.
\begin{equation}
\left\langle \left( \frac{\mathrm{d}\phi }{\mathrm{d}t}\right)
_{a}\right\rangle =\frac{1}{\Gamma },  \label{eq:f11}
\end{equation}%
is turned into energy damage so that no contribution to the current.

\begin{flushleft}
{\large \textbf{4.\hspace*{2mm} Application of the multiple segments model}}
\end{flushleft}

According to the opinions given by Portis\cite{Port}, the CDWs are
pinned by both weak and strong impurities. The strong impurities,
separating the CDW into small segments, are randomly distributed
leading to a distribution of separations $l_{i}$:
\begin{equation}
P(l_{i})=\frac{1}{\left\langle l_{i}\right\rangle }e^{-l_{i}/
\left\langle l_{i}\right\rangle }.  \label{eq:f12}
\end{equation}%
where $\left\langle l_{i}\right\rangle $ is the mean separation
between strong impurities. Each segment is pinned by one strong
impurity such that the pinning potential per length acting by strong
impurities is $\left( \left\langle l_{i}\right\rangle / l_{i}\right)
\sin \phi $. The weak impurities are present in higher concentration
such that treated as a
constant per length on the average and equivalent to a negative field $E_{T}$%
. Thus, the equation for the $l_{i}$th segment of CDW is expected to be of
the form:%
\begin{equation}
\frac{\mathrm{d}^{2}\phi }{\mathrm{d}t^{2}}+\Gamma \frac{\mathrm{d}\phi }{%
\mathrm{d}t}+\frac{\left\langle l_{i}\right\rangle }{l_{i}}\sin \phi =\frac{1%
}{E_{0}}(E-E_{T}),  \label{eq:f13}
\end{equation}%
where the friction coefficient $\Gamma $ arises from thermodynamic damping.
On the analogy of Equation (\ref{eq:f1}), it derives that when the field $E$
is large than a critical field $E_{i}$, this segment will slip, where
\begin{equation}
E_{i}=E_{T}+\frac{l_{i}}{\left\langle l_{i}\right\rangle }E_{0}.
\label{eq:f14}
\end{equation}%
According to equality (\ref{eq:f14}), it derives that, at a given
electric field $E$ there is a critical CDW length $l_{E}$ such that
CDWs longer than $l_{E}$ are slipping and CDWs shorter than $l_{E}$
stay at the equilibrium
points (i.e. foci) where%
\begin{equation}
l_{E}=\left\langle l_{i}\right\rangle \frac{E_{0}}{E-E_{T}}.  \label{eq:f15}
\end{equation}%
As all CDWs connect together, integrating Equation (\ref{eq:f13}),
it
obtains%
\begin{equation}
\int_{0}^{+\infty }\{\frac{\mathrm{d}^{2}\phi }{\mathrm{d}t^{2}}+\Gamma
\frac{\mathrm{d}\phi }{\mathrm{d}t}\}l_{i}P(l_{i})\frac{\mathrm{d}l_{i}}{%
\left\langle l_{i}\right\rangle }=\int_{l_{E}}^{+\infty }\{\frac{l_{i}}{E_{0}%
}(E-E_{T})-\left\langle l_{i}\right\rangle \sin \phi \}P(l_{i})\frac{\mathrm{%
d}l_{i}}{\left\langle l_{i}\right\rangle }.  \label{eq:f16}
\end{equation}%
Completing the integral of Equation (\ref{eq:f16}), it obtains%
\begin{equation}
\frac{\mathrm{d}^{2}\phi }{\mathrm{d}t^{2}}+\Gamma \frac{\mathrm{d}\phi }{%
\mathrm{d}t}+e^{-\frac{E_{0}}{E-E_{T}}}\sin \phi =\frac{1}{E_{0}}%
[E_{0}+(E-E_{T})]e^{-\frac{E_{0}}{E-E_{T}}}.  \label{eq:f17}
\end{equation}%
Let $v_{CDW}$ denotes the velocity by which the ensemble of CDWs
move together, it follows that,
\begin{equation}
v_{CDW}=\frac{\lambda \omega _{0}}{2\pi }\left( \frac{d\phi }{dt}\right)
_{c}.  \label{eq:f17a}
\end{equation}%
The factors $\lambda \omega _{0}(2\pi )^{-1}$ appear above because Equation (%
\ref{eq:f17}) is dimensionless and $\mathrm{d}\phi /\mathrm{d}t$ is a phase
velocity. Comparing Equation (\ref{eq:f17}) with (\ref{eq:f1}), it obtains%
\begin{equation}
v_{CDW}=\frac{\lambda \omega _{0}}{2\pi }\cdot \frac{1}{\Gamma E_{0}}%
(E-E_{T})e^{-\frac{E_{0}}{E-E_{T}}}.  \label{eq:f18}
\end{equation}%
Thus, the conductivity contributed by CDW is%
\begin{equation}
\sigma _{DC}=\frac{nev_{CDW}}{E}=\sigma _{0}(1-\frac{E_{T}}{E})e^{-\frac{%
E_{0}}{E-E_{T}}},  \label{eq:f19}
\end{equation}%
where $\sigma _{0}$ is a measure of the CDW conductivity and $n$ is the
concentration of condensed carriers (i.e. CDW). This formula of conductivity
is the same as the part contributed by CDW in the Fleming's empirical ones%
\cite{Flem}. Meanwhile, the fundamental frequency of narrow-band
noise is
\begin{equation}
f_{1}=\frac{v_{CDW}}{\lambda }=\frac{Q}{2\pi ne}J_{DC},  \label{eq:f20}
\end{equation}%
where $Q$ is the wave vector of CDW, $J_{DC}$ is the direct current density
contributed by CDW. The formula (\ref{eq:f20}), that the fundamental
frequency is proportional to the current, is consistent with the result in
the article\cite{Flem}, which is postulated on experimental ground.

\begin{flushleft}
{\large \textbf{5.\hspace*{2mm} Discussion of Conclusions}}
\end{flushleft}

  The driven damped pendulum is a essential model of nonlinear
science\cite{Hum,Ott}, which develops many manifestations including
chaos and appears in many physical subjects such as the Josephson
junctions\cite{Jacob,Zettl84,Sherwin}. Delicately, the CDW and the
Josephson junction, revealing the chaotic behaviors of conducting
\cite{Shapiro,Hall}, are both well described by the driven damped
pendulum model\cite{MacD}. By a rigorous mathematical analysis are
derived these conclusions of the pendulum equation applied in
\cite{Arxiv}, which provide us a more direct relation between the
pendulum model and the behavior of the CDW. Such a simple model of
pendulums indicates us an interesting relations of these physical
subjects.

\vspace{0.4cm} \noindent{\large \textbf{Acknowledgments}}

The author thanks the referee for his/her constructive and useful
recommendations, thanks Professor L. Zhao for his help on LaTex software.



\end{document}